\documentclass[aps,prd,twocolumn,superscriptaddress,amsmath,showpacs,floatfix,nofootinbib]{revtex4}
\usepackage{graphicx}
\usepackage{dcolumn}
\usepackage{bm}
\usepackage{natbib}
\usepackage{multirow}

\newcommand{\nrun}{n_\mathrm{run}}
\newcommand{\neff}{N_\mathrm{eff}}
\newcommand{\Mpc}{{\rm ~Mpc}}

\begin{document}

\title{Cosmological data and indications for new physics}

\author{Micol Benetti}
\affiliation{Physics Department and ICRA, Universit\`a di Roma 
	``La Sapienza'', Ple.\ Aldo Moro 2, 00185, Rome, Italy}
\affiliation{Physics Department and INFN, Universit\`a di Roma ``La Sapienza'', Ple Aldo Moro 2, 00185, Rome, Italy}

\author{Martina Gerbino}
\affiliation{Physics Department and INFN, Universit\`a di Roma ``La Sapienza'', Ple Aldo Moro 2, 00185, Rome, Italy}

\author{William H.\ Kinney} \email{whkinney@buffalo.edu}
\affiliation{Department of Physics, University at Buffalo, the State University of New York, Buffalo, NY 14260-1500}

\author{Edward W.\ Kolb} \email{Rocky.Kolb@uchicago.edu}
\affiliation{Department of Astronomy and Astrophysics, Enrico Fermi Institute, and Kavli Institute for Cosmological Physics, University of Chicago, Chicago, Illinois \ 60637-1433}

\author {Massimiliano Lattanzi} \email{lattanzi@fe.infn.it}
\affiliation{Dipartimento di Fisica e Science della Terra, Universit\`a di Ferrara and INFN, sezione di Ferrara, Polo Scientifico e Tecnologico - Edificio C Via Saragat, 1, I-44122 Ferrara Italy}

\author{Alessandro Melchiorri}
\affiliation{Physics Department and INFN, Universit\`a di Roma ``La Sapienza'', Ple Aldo Moro 2, 00185, Rome, Italy}

\author{Luca Pagano}
\affiliation{Physics Department and INFN, Universit\`a di Roma ``La Sapienza'', Ple Aldo Moro 2, 00185, Rome, Italy}

\author{Antonio Riotto} \email{antonio.riotto@unige.ch}
\affiliation{Department of Theoretical Physics and Center for Astroparticle Physics (CAP) 24 quai E. Ansermet, CH-1211 Geneva 4, Switzerland}

\begin{abstract}
Data from the Atacama Cosmology Telescope (ACT) and the South Pole Telescope (SPT), combined with the nine-year data release from the WMAP satellite, provide very precise measurements of the cosmic microwave background (CMB) angular anisotropies down to very small angular scales.  Augmented with measurements from Baryonic Acoustic Oscillations surveys and determinations of the Hubble constant, we investigate whether there are indications for new physics beyond a Harrison-Zel'dovich model for primordial perturbations and the standard number of relativistic degrees of freedom at primordial recombination. All combinations of datasets point to physics beyond the minimal Harrison-Zel'dovich model in the form of either a scalar spectral index different from unity or additional relativistic degrees of freedom at recombination (\textit{e.g.}, additional light neutrinos).  Beyond that, the extended datasets including either ACT or SPT  provide very different indications: while the extended-ACT (eACT) dataset is perfectly consistent with the predictions of standard slow-roll inflation, the extended-SPT (eSPT) dataset prefers a non-power-law scalar spectral index with a very large variation with scale of the spectral index. Both eACT and eSPT favor additional light degrees of freedom. eACT is consistent with zero neutrino masses, while eSPT favors nonzero neutrino masses at more than 95\% confidence.
\end{abstract}
 

\maketitle

\section{Introduction} \label {sec:intro}

A wide variety of observations of Cosmic Microwave Background (CMB) and Large Scale Structure (LSS) power spectra over the last decade have indicated that cosmic structures originated from seed fluctuations in the very early universe. The leading theory explaining the origin of the cosmological seed perturbations is cosmic inflation \cite{lrreview}, a period of accelerated expansion at very early times. During the inflationary stage, microscopic quantum fluctuations were stretched to macroscopic scales to provide both the initial seeds for the primordial density perturbations and tensor (gravitational-wave) fluctuations \cite{Starobinsky:1979ty,muk81,Guth:1982ec,Hawking:1982cz,bardeen83}.  Despite the simplicity of the inflationary paradigm, the exact mechanism by which cosmological perturbations are generated is not yet established. 

In the standard slow-roll inflationary scenario associated with the dynamics of a single scalar field (the inflaton), density perturbations are due to fluctuations of the inflaton itself as it slowly rolls down along its potential. In the simplest case, fluctuations are of the adiabatic type, namely they are sourced by the degree of freedom that is dominating the energy density during inflation (the inflaton).  In other mechanisms for the generation of perturbations, {\it e.g.,} the curvaton mechanism \cite{curvaton}, the final adiabatic perturbations are produced from an initial isocurvature mode associated with quantum fluctuations of a light scalar degree of freedom (other than the inflaton), whose energy density is negligible during inflation. The isocurvature perturbations are then transformed into adiabatic perturbations when the the extra scalar degree of freedom (the curvaton) decays into radiation after the end of inflation.  A precise measurement of the spectral index, $n_S$, of the scalar perturbations together with a detection of  gravity-wave signals in CMB anisotropies through its $B$-mode polarization will provide a strong hint in favor of single-field models of inflation. Indeed, alternative mechanisms predict an amplitude of gravity waves far too small to be detectable by experiments aimed at observing the $B$-mode of the CMB polarization. 

While inflation is the leading candidate model for the generation of primordial perturbations, in the words of Ref.\ \cite{Martinec:2012bv}: ``Inflation is at the same time a spectacular phenomenological success, and an enduring theoretical challenge.'' The phenomenological success is that inflation is a simple model for the generation of seed perturbations.  The theoretical challenge is to understand how inflation is embedded in a broader theory or model of fundamental physics. Detailed examination of the CMB perturbations are a possible way to discriminate among inflation models \cite{Dodelson:1997hr}, perhaps even leading to a reconstruction of the inflaton potential \cite{Lidsey:1995np}.

Another application of CMB data is to search for evidence of ``new'' physics, like additional relativistic degrees of freedom at recombination \cite{knox11}, neutrino masses \cite{Lesgourgues:2006nd}, early dark energy \cite{Calabrese:2011hg}, modified gravity \cite{DiValentino:2012yg}, or variation of fundamental constants like the fine-structure constant \cite{Menegoni:2012tq} or the gravitational constant \cite{Galli:2009pr}.

The goal of this paper is to examine whether existing cosmological datasets can provide evidence for the dynamics of the inflaton field during inflation or evidence for new ``neutrino'' physics.  Evidence for the dynamics of the inflaton field during inflation would be a departure from the Harrison--Zel'dovich (HZ) model (a scalar spectral index of unity and no tensor perturbations).\footnote{We note that a scalar spectral index of unity is possible within slow-roll inflation \cite{Vallinotto:2003vf,Starobinsky:2005ab}.}  The departure from the HZ model could take the form of a scalar spectral index different than unity, a ``running'' (a scale-dependence) of the scalar spectral index, or evidence for tensor modes.

The type of new ``neutrino'' physics we model would be a mass for neutrinos or additional relativistic degrees of freedom contributing to the expansion rate around the time of recombination.\footnote{Although we parameterize the additional relativistic degrees of freedom as a contribution to the effective number of light neutrinos, $\neff$, of course the new relativistic species need not be neutrinos.}

Our analysis will include CMB data from the nine-year data release of the Wilkinson Microwave Anisotropy Probe (WMAP9) \cite{wmap9}, the South Pole Telescope (SPT) \cite{spt2013}, and the Atacama Cosmology Telescope (ACT) \cite{act2013}, including measurements up to a maximum multipole number of $l_{\rm max}\simeq 3000$.  We will also include information from measurements of baryonic acoustic oscillations (BAO) from galaxy surveys in the form of three datasets: data release 7 (SDSS-DR7) \cite{padmanabhan/etal:2012} and data release 9 (SDSS-DR9) \cite{anderson/etal:2012} from the Sloan Digital Sky Survey, and the WiggleZ project \cite{blake/etal:2012}.   We will also use data on the Hubble constant \cite{hst}.

This study has two motivations. On one hand, the Planck collaboration \cite{planck} will release soon their first flow of data regarding the CMB anisotropies, and therefore it is timely to have a state-of-the-art pre-Planck assessment of slow-roll inflation. On the other hand, we wish to answer three basic questions:
\begin{enumerate}
\item Is the simple Harrison-Zel'dovich model compatible with  current cosmological datasets, or is there support for a more complicated perturbation spectrum?
\item Is standard neutrino physics consistent with current cosmological datasets, or is there support for new neutrino physics in the form of a neutrino mass in excess of a few tenths of an electron volt or a change in the effective number of light neutrinos?
\item In the event that there is support for physics beyond the HZ spectrum and standard neutrinos, can one tell whether the data provides information about the primordial perturbation spectrum or the neutrino sector?
\end{enumerate}

As we will see, all combinations of current datasets point to physics beyond the minimal HZ model with standard neutrino physics. In particular, allowing for either a scalar spectral index different from unity or an additional number of relativistic degrees of freedom, produces a significant increase in the goodness of fit with respect to the minimal HZ model. Information beyond that depends on the dataset used.

For a dataset with ACT (and not SPT), there is no significant increase in the goodness of fit by increasing the complexity of the perturbation spectra by allowing for running of the scalar perturbations and/or a tensor component.  Concerning neutrino physics, there is no significant further increase in the goodness of fit by allowing for a nonzero neutrino mass.  For the dataset with ACT, we can only say that a model with a scalar spectral index different than unity is a much better fit than the HZ model, and a model with a non-standard number of neutrinos is also a much better fit.  In the sections below we will quantify these statements.

The situation is much different if we examine a dataset with SPT (and not ACT).\footnote{Ref.\ \cite{Calabrese:2013jyk} concluded that datasets including either ACT or SPT give a consistent picture for cosmological parameters, as long as HZ$+n_S$ and HZ$+n_s+\neff$ are concerned. We find however that the answer to the question of whether cosmological data points to physics beyond an HZ $+n_S$ model strongly depends on the choice of whether the dataset includes ACT or SPT.} There are a significant increases in goodness of fit allowing either a running of the scalar spectral index (and not much increase in goodness of fit just by allowing a tensor component) on top of the HZ+$n_s$ model, or a nonzero neutrino mass to the HZ  model. Again, in the sections below we will quantify these statements.

Therefore, we conclude that a cosmological dataset including SPT suggests either a more complex perturbation spectrum than simply a scalar spectral index different than unity, or some other new physics such as a modification of the number of relativistic degrees of freedom.  The data do not prefer one approach over the other.

The papers is organized as follows:  In the next section we review the pertinent features of slow-roll inflation.  In Sec.\ \ref{sec:neutrinos} we review how neutrinos (or other light species) affect the CMB anisotropies.  In Sec.\ \ref{sec:method} we discuss our data analysis method and the datasets examined.  Section \ref{sec:extensions} presents our results for cosmological parameters and the maximum likelihood for various models.  In Sec.\ \ref{sec:quovadis} we discuss implications for physics beyond the HZ model for neutrino physics and for inflation. Section \ref{sec:conclusions} contains our conclusions.

\section{Slow-roll inflation and CMB anisotropies} \label{sec:slowroll}

As mentioned in the introduction, we will work under the hypothesis that the adiabatic perturbations originated within the single-field, slow-roll framework of inflation. It should be kept in mind that if future experiments do not detect isocurvature modes or large non-Gaussianity it will not be possible to distinguish directly the inflaton contribution from the, {\it e.g.,} curvaton contribution, see Ref.\ \cite{az}. On the other hand, a detection of a significant amount of tensor modes through  CMB anisotropies will disfavor curvaton-like models as they tend to generate a negligible tensor contribution. 
 
Within the single-field slow-roll paradigm, many specific models for inflation have been proposed. We limit ourselves here to models with ``normal'' gravity ({\em i.e.,} general relativity) and a single order parameter for the vacuum, described by a canonical scalar field $\phi$, the inflaton, with Lagrangian
\begin{equation}
{\mathcal L} = \frac{1}{2} g^{\mu\nu} \partial_\mu \phi \partial_\nu \phi - V\left(\phi\right).
\end{equation}
The equations of motion for the spacetime are given by the Friedmann Equations,  which for a homogeneous field $\phi$ are
\begin{eqnarray}
\label{eq:Friedmann}
H^2 = \left(\frac{\dot a}{a}\right)^2 &=& \frac{8 \pi}{3 m_{\rm Pl}^2}\left[\frac{1}{2} \dot\phi^2 + V\left(\phi\right)\right],\cr
\left(\frac{\ddot a}{a}\right) &=& - \frac{4 \pi}{3m_{\rm Pl}^2} \left[\dot\phi^2 - V\left(\phi\right)\right]. 
\end{eqnarray}
The equation of motion for the field $\phi$ is
\begin{equation}
\label{eq:phieom}
\ddot \phi + 3 H \dot\phi + V'\left(\phi\right) = 0.
\end{equation}
We have assumed a flat Friedmann-Robertson-Walker metric $g_{\mu \nu} = {\rm diag}(1, -a^2, -a^2 -a^2)$, where $a(t)$ is the scale factor of the universe.  Inflation is defined to be a period of accelerated expansion, $\ddot a >0$.  If the field evolution is monotonic in time, we can write the scale factor $a\left(\phi\right)$ and Hubble parameter $H\left(\phi\right)$ as functions of the field $\phi$ rather than time, {\it i.e.,} we define all of our physical parameters along the trajectory in phase space ${\dot\phi}\left(\phi\right)$ corresponding to the classical solution to the equations of motion. Equations (\ref{eq:Friedmann}) and (\ref{eq:phieom}) can then be re-written exactly in the Hamilton-Jacobi form
\begin{eqnarray} 
& &\dot\phi =  -\frac{m_{\rm Pl}^2}{4 \pi} H'(\phi),\cr
& & \left[H'(\phi)\right]^2 - \frac{12 \pi}{m_{\rm Pl}^2}
H^2(\phi) = - \frac{32 \pi^2}{m_{\rm Pl}^4}
V(\phi).
\label{eqbasichjequations}
\end{eqnarray}
These are completely equivalent to the second-order equation of motion. The second of the above equations is referred to as the Hamilton-Jacobi equation, and can be written in the useful form
\begin{equation} 
H^2(\phi) \left[1 - \frac{1}{3}
\epsilon(\phi)\right] =  \left(\frac{8 \pi}{3 m_{\rm Pl}^2}\right) V(\phi),
\label{eqhubblehamiltonjacobi}
\end{equation}
where $\epsilon$ is defined to be
\begin{equation}
\epsilon(\phi) \equiv \frac{m_{\rm Pl}^2}{4 \pi} \left(\frac{H'(\phi)}{ H(\phi)}\right)^2.\label{eqdefofepsilon}
\end{equation}
The physical meaning of $\epsilon(\phi)$ can be seen by expressing $\ddot a$ in Eq.\ 
(\ref{eq:Friedmann}) as
\begin{equation}
\left(\frac{\ddot{a}}{a}\right) = H^2 (\phi) \left[1 -
\epsilon(\phi)\right],
\end{equation}
so that the condition for inflation, $(\ddot a / a) > 0$, is equivalent to $\epsilon < 1$. The scale factor is given by
\begin{equation}
a \propto e^{N} = \exp\left[\int_{t_0}^{t}{H\,dt}\right],
\end{equation}
where the number of $e$-folds $N$ is
\begin{equation}
N \equiv \int_{t}^{t_e}{H\,dt} = \int_{\phi}^{\phi_e}{\frac{H}{\dot\phi}\,d\phi} = \frac{2 \sqrt{\pi}}{m_{\rm Pl}}
\int_{\phi_e}^{\phi}\frac{d\phi}{\sqrt{\epsilon(\phi)}}.\label{eqdefofN}
\end{equation}

Most simple inflation models satisfy the slow-roll approximation, which is the assumption that the evolution of the field is dominated by the drag from the cosmological expansion, so that $\ddot\phi \simeq 0$ and  $\dot \phi \simeq -V'/3 H$. The equation of state of the scalar field is dominated by the potential, so that $p \simeq -\rho$, and the expansion rate is approximately $H^2 \simeq 8 \pi V(\phi)/ 3 m_{\rm Pl}^2$. The slow roll approximation is consistent if both the slope and curvature of the potential are small, $V',\ V'' \ll V$ (in units of the Planck mass $m_{\rm Pl}$). In this case the parameter    $\epsilon$ can be expressed in terms of the potential as
\begin{equation}
\epsilon \equiv \frac{m_{\rm Pl}^2}{4 \pi} \left(\frac{H'\left(\phi\right)}{H\left(\phi\right)}\right)^2 
\simeq \frac{m_{\rm Pl}^2}{16 \pi}
\left(\frac{V'\left(\phi\right)}{V\left(\phi\right)}\right)^2.
\end{equation}
We will also define a second ``slow-roll parameter'' $\eta$ by
\begin{eqnarray}
\eta\left(\phi\right) &\equiv& \frac{m_{\rm Pl}^2}{4 \pi} 
\left(\frac{H''\left(\phi\right)}{H\left(\phi\right)}\right)\cr
&\simeq& \frac{m_{\rm Pl}^2}{8 \pi}
\left[\frac{V''\left(\phi\right)}{V\left(\phi\right)} - \frac{1}{2}
\left(\frac{V'\left(\phi\right)}{V\left(\phi\right)}\right)^2\right].
\end{eqnarray}
Slow roll is then a consistent approximation for $\epsilon,\ \eta \ll 1$. 

Perturbations created during inflation are of two types: scalar (or curvature) perturbations, which couple to the stress-energy of matter in the universe and form the ``seeds'' for structure formation, and tensor, or gravitational-wave perturbations, which do not couple to matter.  Both scalar and tensor perturbations contribute to CMB anisotropies. Scalar fluctuations can also be interpreted as fluctuations in the density of the matter in the universe. Scalar fluctuations can be quantitatively characterized by the comoving curvature perturbation $P_{\cal R}$. As long as  slow roll is attained, the curvature (scalar) perturbation at horizon crossing  can be shown to be \cite{lrreview} 
\begin{equation} 
P_{\cal R}^{1/2}\left(k\right) =
\left(\frac{H^2}{2 \pi \dot \phi}\right)_{k = a H} =    
\left [\frac{H}{m_{\rm Pl} } \frac{1}{\sqrt{\pi \epsilon}}\right]_{k = a H}. 
\end{equation} 
The fluctuation power spectrum is, in general, a function of wavenumber $k$, and is evaluated when a given mode crosses outside the horizon during inflation, $k = a H$. Outside the horizon, modes do not evolve, so the amplitude of the mode when it crosses back inside the horizon during a later radiation- or matter-dominated epoch is just its value when it left the horizon during inflation.  Instead of specifying the fluctuation amplitude directly as a function of $k$, it is convenient to specify it as a function of the number of $e$-folds $N$ before the end of inflation at which a mode crossed outside the horizon. 

The scalar spectral index $n_S$ for $P_{\cal R}$ is defined by
\begin{equation}
n_S - 1 \equiv \frac{d\ln P_{\cal R}}{d\ln k},
\end{equation}
so that a scale-invariant spectrum, in which modes have constant amplitude at horizon crossing, is characterized by $n_S = 1$. 

To lowest order in slow roll, the power spectrum of tensor fluctuation modes and the corresponding tensor  spectral index is given by \cite{lrreview}
\begin{eqnarray}
P_T^{1/2}\left(k_N\right) & = & \left[\frac{4 H}{m_{\rm Pl} \sqrt{\pi}}
\right]_{k=aH}, \nonumber \\
n_T & \equiv & \frac{d\ln P_T}{d\ln k}.
\end{eqnarray}
The ratio of tensor-to-scalar modes is then $ P_T/P_{\cal R} = 16 \epsilon$, so that tensor modes are negligible for $\epsilon \ll 1$. In the limit of slow roll, the spectral indices $n_S$ and $n_T$ vary slowly or not at all with scale.  We can write the spectral indices $n_S$ and $n_T$ to lowest order in terms of the slow-roll parameters $\epsilon$ and $\eta$ as
\begin{eqnarray}
n_S & \simeq & 1 - 4 \epsilon + 2 \eta,\nonumber \\
n_T & \simeq& - 2 \epsilon.
\end{eqnarray}

The tensor/scalar ratio is frequently expressed as a quantity $r$, which is conventionally normalized as
\begin{equation}
r \equiv 16 \epsilon = \frac{P_T}{P_{\cal R}} .
\end{equation}
The tensor spectral index is not an independent parameter, but is proportional to the tensor/scalar ratio, given to lowest order in slow roll by $ n_T \simeq - 2 \epsilon = - r/8$. A given inflation model can therefore be described to lowest order in slow roll by three independent parameters: $P_{\cal R}$, $P_T$, and $n_S$. 

Deviations from a simple power-law spectrum of perturbations are higher order in the slow-roll parameters, and thus serve as a test of the consistency of the slow-roll approximation. Scale dependence in the observables corresponds to scale dependence in the associated slow-roll parameter, and can be quantified in terms of the infinite hierarchy of inflationary flow equations \cite{Kinney:2002qn},
\begin{eqnarray}
\frac{d \epsilon}{d N} &=& 2 \epsilon \left(\eta - \epsilon\right), \cr
\frac{d \eta}{d N} &=&  {}^2 \lambda - \epsilon \eta, \cr
\vdots \cr  
\frac{d {}^\ell \lambda} {d N} &=& \left[(\ell-1) \eta - \ell \epsilon \right] {}^ \ell \lambda + {} ^ {(\ell + 1)} \lambda. 
\label{eq:flowequations}
\end{eqnarray}
The higher-order flow parameters are defined by
\begin{eqnarray}
\label{eq:definflationparamshierarchy}
\epsilon &\equiv& {2 M_{P}^2 } \left(\frac{H'(\phi)}{H(\phi)}\right)^2, \cr 
\eta &\equiv& {2 M_{P}^2 } \frac{H''(\phi)}{H(\phi)}, \cr
{}^2 \lambda  &\equiv& {4 M_{P}^4}  \frac{H'(\phi) H'''(\phi)}{H^2(\phi)}, \cr 
\vdots \cr   
{}^ \ell \lambda &\equiv& {\left(2 M_{P}^2\right) ^ \ell} \frac{H'(\phi)^{\left(\ell-1\right)}}{H(\phi)^ \ell} \frac{d^{\left(\ell +1\right)} H(\phi)}{d \phi ^ {\left(\ell +1\right)}},
\end{eqnarray}
where the prime denotes derivatives with respect to scalar field $\phi$. It is then straightforward to calculate the scale-dependence of the spectral index by relating the wavenumber $k$ to the number of $e$-folds $N$,
\begin{eqnarray}
\frac{d n}{d \ln k} & \equiv & \nrun = - \frac{1}{1 - \epsilon} \frac{d}{dN} \left(2 \eta - 4 \epsilon\right)\cr
& = &10 \epsilon \eta - 8 \epsilon^2 - 2\left({}^2\lambda\right) + {\mathcal O}(\epsilon^3) + \cdots.
\label{eq:dndlnk}
\end{eqnarray}
Since the running depends on higher-order flow parameters than the spectral index itself, it is an independent parameter,  even in slow-roll inflation models. In typical single-field inflation models, the running of the spectral index is negligible, so a detection of scale dependence in the spectral index would rule out a large class of viable single-field inflation models, and would therefore be a powerful probe of inflationary physics.

\section{Neutrinos and CMB anisotropies} \label {sec:neutrinos}

In what follows we we examine the possibility of new neutrino physics as an alternative to extending the complexity of primordial perturbations.

One direction for new neutrino physics is a change in the effective number of relativistic degrees of freedom, $\neff$, that defines the physical energy density in relativistic particles $\rho_\mathrm{rad}$, defined by 
\begin{equation}
     \rho_{\rm rad}=\left[ 1+ \frac{7}{8} \left(\frac{4}{11}\right)^{4/3} 
          \neff \right ]\rho{_\gamma} \ ,
\end{equation}
where $\rho_\gamma$ is the energy density of the CMB photons and $\neff$ is the effective number of light neutrino species. In the standard scenario, assuming three active massless neutrino species  with standard electroweak interactions and the present CMB temperature of $T_{\gamma}=2.726K$ (see, e.g., Ref.\ \cite{fixsen}), the expected value is $\neff=3.046$. This is slightly larger than $3$ because of non-instantaneous neutrino decoupling (see, e.g., Ref.\ \cite{mangano3046}). As mentioned previously, any new species that is relativistic around recombination will contribute to $\neff$, whether it is a neutrino species or not.  The exact contribution of a new relativistic species will depend on the number of spin degrees of freedom, whether the new species is a boson or fermion, and the temperature of decoupling of the new species.

We also consider the possibility of a mass for one or more of the three known active neutrino species.  The present contribution to the overall energy density is given by
\begin{equation}
     \Omega_{\nu}h^2= \Sigma_{i=1,2,3} \ \ \frac{m_i}{92.5\ \mathrm{eV}},
\end{equation}
where $m_i$ are the masses of the three neutrino mass eigenstates.

A change in neutrino physics can have important implications for interpretation of inflationary parameters from CMB anisotropies, see Refs.\ \cite{julienne,bowen,archidiacono}.  For example, varying $\neff$ can have an impact on determination of $n_S$ and its running, since it changes both the position of the CMB peaks in the angular spectrum and the structure of the ``damping tail'' at very large multipoles (see Ref.\ \cite{knox11}).  In general, a higher $\neff$ can put higher values of $n_S$ in better agreement with the data, {\textit i.e.,} there is a positive correlation between the two parameters.

Masses for neutrinos also have important implications for interpretation of inflationary parameters from CMB anisotropies.  Massive neutrinos damp the dark-matter fluctuations on scales below the horizon when they become nonrelativistic (see {\textit e.g.,} \cite{hu1998}). Neutrinos with masses $m_\nu \alt 0.3$ eV are relativistic at recombination and affect the CMB anisotropy mainly through gravitational lensing, while neutrinos with larger masses slightly increase the CMB small-scale anisotropy by damping the gravitational potential at recombination. The final result is a small anti-correlation with $n_S$, {\textit i.e.,} larger neutrino masses shift the constraints on $n_S$ to smaller values.

\section{Data Analysis Method} \label{sec:method}

The analysis method we adopt is based on the publicly available Monte Carlo Markov Chain (MCMC) package \texttt{cosmomc} \cite{Lewis2002} with a convergence diagnostic done through the Gelman and Rubin statistic.

We sample the following four-dimensional standard set of cosmological parameters, adopting flat priors on them: the baryon and cold dark matter densities $\Omega_ b$ and $\Omega_c$, the angular size of the sound horizon at decoupling $\theta$, and the optical depth to reionization $\tau$.

As discussed in a separate section, we will also vary the relativistic number of degrees of freedom parameter $\neff$ and the total neutrino mass $\Sigma m_{\nu}$. The standard three-neutrino framework predicts $\neff=3.046$, while oscillation neutrino experiments place a lower bound $\Sigma m_{\nu} > 0.05$ eV \cite{Fogli:2006zz}. 

For the inflationary parameters we consider the scalar spectral index $n_S$ and its running $\nrun$, the overall normalization of the spectrum $A_S$ at $k=0.002\Mpc^{-1}$ and the amplitude of the tensor modes relative to the scalar, $r=A_T/A_S$, again at $k=0.002\Mpc^{-1}$.

We consider purely adiabatic initial conditions and we impose spatial flatness.

We analyze the following set of CMB data: WMAP9 \cite{wmap9}, SPT \cite{spt2013}, and ACT \cite{act2013}, including measurements up to a maximum multipole number of $l_{\rm max}\simeq 3000$.  For all these experiments we make use of the publicly available codes and data. For the ACT experiment we use the ``lite'' version of the likelihood \cite{dunkleyact}. 

We also consider the effect of including additional datasets to the basic datasets just described.  Consistently with the measurements of HST \cite{hst}, we consider a Gaussian prior on the Hubble constant $H_0=73.8\pm2.4 \,\mathrm{km}\,\mathrm{s}^{-1}\,\mathrm{Mpc}^{-1}$. We also include information from measurements of baryonic acoustic oscillations (BAO) from galaxy surveys. Here, we follow the approach presented in Ref.\ \cite{wmap9} combining three datasets: SDSS-DR7 \cite{padmanabhan/etal:2012}, SDSS-DR9 \cite{anderson/etal:2012} and WiggleZ \cite{blake/etal:2012}.

Since, as we see in the next section, the ACT and the SPT datasets are providing significantly different conclusions on inflationary parameters, we will include them separately.  In what follows we will consider two combinations of datasets.  We refer to an analysis using the WMAP9 $+$ ACT $+$ HST $+$ BAO datasets as the ``extended ACT'' (eACT) dataset and to an analysis with the WMAP9 $+$ SPT $+$ HST $+$ BAO datasets as the ``extended SPT'' (eSPT) dataset.

We use the Markov chains obtained from CosmoMC to reconstruct the posterior distributions of each of the model parameters. In the tables, we present our results in the form of the 68\% credible interval for each parameter, {\textit i.e.,} the interval symmetric around the mean containing 68\% of the total posterior probability. We make an exception to this rule in those cases where the posterior probability is not vanishingly small at the edge of the prior range; this happens in particular around $r=0$ and $\sum m_\nu = 0$. In this case we adopt the following rule: if the maximum of the posterior distribution  is clearly distinguished from zero, we quote the 68\% interval as above; otherwise, we quote a 95\% upper limit. 

We also use our Markov chains to recover the maximum likelihood (\textit{i.e.}, minimum $\chi^2$) parameter values. We use the minimum $\chi^2$ values estimated from the chains to perform an approximate model comparison by computing the likelihood ratio (actually, equivalently, the difference in $\chi^2$) between models. As a rule of thumb, 
given two models ${\mathcal M}_1$ and ${\mathcal M}_2$, where the latter reduces to the former for a particular choice of parameter values (in which case the two models are said to be ``nested''), we say that the data show preference for ${\mathcal M_2}$ over ${\mathcal M_1}$ when the absolute value of $\Delta\chi^2 \equiv \chi^2_\mathrm{min}({\mathcal M}_2) -\chi^2_\mathrm{min}({\mathcal M}_1)$ is larger than the number of additional parameters in the extended model.

We note however that MCMC methods are usually optimized to sample the full posterior distribution around the region of maximum probablity, and not to recover the exact value and position of the maximum likelihood. The precision to be associated with our estimate of the minimum $\chi^2$ can be evaluated by computing the probability of finding in the chains a sample having a $\chi^2$ within $\delta(\chi^2)$ from the actual minimum (see e.g. \cite{Hamann:2011hu}). This probability depends on the dimensionality of the parameter space and on the number of independent samples in the chains. We let our chains run until we can claim a 95\% probability of having found the best-fit model with an uncertainty $\delta(\chi^2) \le 1$. 

\section{Extensions of the HZ model and eACT and eSPT \label{sec:extensions}} 

As a first step in our analysis we evaluate the compatibility of current cosmological datasets eSPT and eACT with a simple reference model, which we choose to be the Harrison-Zel'dovich (HZ) model with $n_S=1$, $r=0$, $\nrun=0$, $m_\nu=0$, and $\neff=3.046$.  We then consider extensions of this model involving more complex perturbation spectra, with various combinations of $n_S \neq 1$, $r>0$, and $\nrun\neq0$.  Then we examine extensions of the HZ model with nonstandard neutrino physics with combinations of $m_\nu\neq0$ and/or $\neff\neq 3.046$.

\subsection{Extensions of the perturbation sector \label{sec:infquovadis}}

\renewcommand*\arraystretch{1.5}
\begin{table*}[htb!]
\caption{Augmenting the minimal Harrison-Zel'dovich cosmological model through inflationary parameters. Listed are posterior means for the cosmological parameters from the indicated datasets (errors refer to 68\% confidence intervals, unless otherwise stated).}
\begin{ruledtabular}
\label{tab:Stew_inf}
\footnotetext [1] {km s$^{-1}$ Mpc$^{-1}$}
\footnotetext[2]{When comparing to the $\chi^2$ values reported e.g. in the WMAP9 paper \cite{wmap9}, it should be taken into account that we use a pixel based likelihood at low $l$s instead than the Gibbs-based likelihood.}
\footnotetext [3] {$\Delta\chi^2\equiv \left(- 2\log\mathcal{L}\right)-\left(- 2\log\mathcal{L}_{HZ}\right)$}
\footnotetext [4] {95\% c.l.}
\begin{tabular}{c|c|c|cccc}
\multirow{2}{*}{Dataset} & \multirow{2}{*}{Parameter} & Reference Model \phantom{X}& \multicolumn{4}{c}{Inflation-Motivated Extensions} \\
& & HZ & HZ $+n_S$ & HZ $+n_S+r$ & HZ $+n_S+\nrun$ & HZ $+n_S+r+\nrun$ \\ \hline
\multirow{10}{*}{eSPT} & $100\,\Omega_b h^2$	
& $2.331\pm0.025$ &$2.225\pm0.032$&$2.228\pm0.032$  &$2.236\pm0.031$  &$2.272\pm0.036$ \\
& $\Omega_c h^2$        
& $0.1148\pm0.0017$&$0.1167\pm0.0018$ &$0.1166\pm0.0018$ &$0.1180\pm0.0019$&$0.1178\pm0.0018$ \\
& $100\, \theta$          
& $1.0430\pm0.0009$ &$1.0419\pm0.0009$ &$1.0419\pm0.0010$ &$1.0422\pm0.0009$ &$1.0424\pm0.0009$ \\       
& $\log[10^{10}A_S]$ \phantom{X}
& $ 3.12\pm0.03$ &$3.21\pm0.03$ &$3.20\pm0.03$ &$3.14\pm0.04$ &$3.04\pm0.07$ \\
& $\tau$
& $0.096\pm0.013$&$0.078\pm0.012$   &$0.077\pm0.012$ &$0.090\pm0.014$	&$0.095\pm0.015$ \\
& $n_S$
& $\equiv1$&$0.959\pm0.008$   &$0.962\pm0.008$ &$1.037\pm0.029$	&$1.107\pm0.045$ \\
& $r$
& $\equiv0$&$\equiv0$ &$< 0.12$  \ \ $^{(d)}$ &$\equiv0$ &$0.28\pm0.16$ \\
& $\nrun$			
& $\equiv0$&$\equiv0$ &$\equiv0$ &$-0.029\pm0.011$&$-0.051\pm0.015$\\
& $H_0$ \ \ $^{(a)}$
& $71.33\pm0.65$ &$69.33\pm0.74$ 	&$69.42\pm0.76$	&$69.08\pm0.76$	&$69.51\pm0.78$\\
& $- 2\log\mathcal{L}$ \ \ $^{(b)}$	       		
& $7653.4 $ &$7624.7$ &$7625.6$  & $7616.8$   &$7615.9$\\
& $\Delta\chi^2$\ \ $^{(c)}$
& $\equiv0$ & $-28.7$ & $-27.8$ & $-36.6$ & $-37.5$ \\
\hline
\multirow{10}{*}{eACT} & $100\,\Omega_b h^2$	
& $2.356\pm0.027$&$2.282\pm0.035$&$2.290\pm0.037$&$2.283\pm0.035$&$2.302\pm0.038$\\
& $\Omega_{c} h^2$			
& $0.1163\pm0.0021$&$0.1165\pm0.0021$&$0.1162\pm0.0021$&$0.1166\pm0.0021$&$0.1167\pm0.0022$\\
& $100\, \theta$                		
& $1.0416\pm0.0016$&$1.0399\pm0.0018$&$1.0399\pm0.0017$&$1.0400\pm0.0017$&$1.0403\pm0.0018$\\
& $\log[10^{10} A_S]$ \phantom{X}		
& $3.14\pm0.03$&$3.19\pm0.03$&$3.18\pm0.03$&$3.19\pm0.04$&$3.13\pm0.05$\\
& $\tau$ 		                  		
& $0.102\pm0.014$&$0.090\pm0.014$&$0.089\pm0.013$&$0.092\pm0.015$&$0.094\pm0.015$\\
& $n_S$		
& $\equiv1$&$0.971\pm0.009$&$0.976\pm0.009$&$0.978\pm0.031$&$1.016\pm0.042$\\
& $r$		
& $\equiv0$&$\equiv0$&$<0.18$ \ \ $^{(d)}$ &$\equiv0$&$<0.34$ \ \ $^{(d)}$\\
& $\nrun$			
& $\equiv0$&$\equiv0$&$\equiv0$&$-0.003\pm0.011$&$-0.014\pm0.014$\\
& $H_0$ \ \ $^{(a)}$
& $70.50\pm0.71$ &$69.24\pm0.83$&$69.43\pm0.83$&$69.24\pm0.81$&$69.47\pm0.83$\\
& $- 2\log\mathcal{L}$ $^{(b)}$	       	
& $7617.9$&$7608.2$&$7608.4$&$7608.3$&$7608.7$\\
& $\Delta\chi^2$\ \ $^{(c)}$
& $\equiv0$ & $-9.7$ & $-9.5$ & $-9.6$ & $-9.2$ \\
\end{tabular}
\end{ruledtabular}
\end{table*}

The results of our analysis with regard to perturbation spectra is reported in Table \ref{tab:Stew_inf}.  As stated in the previous section, we analyze the eACT and eSPT datasets and we consider different cases for primordial perturbations and compare them with the reference HZ model.  In all models analyzed in this section we assume massless neutrinos and $\neff=3.046$.

As we can see from the table, both for the eACT and eSPT datasets, models with $n_S\neq1$ are highly favored over the HZ reference model.  

\begin{figure*}
\begin{center}
\includegraphics[width=0.9\linewidth,keepaspectratio]{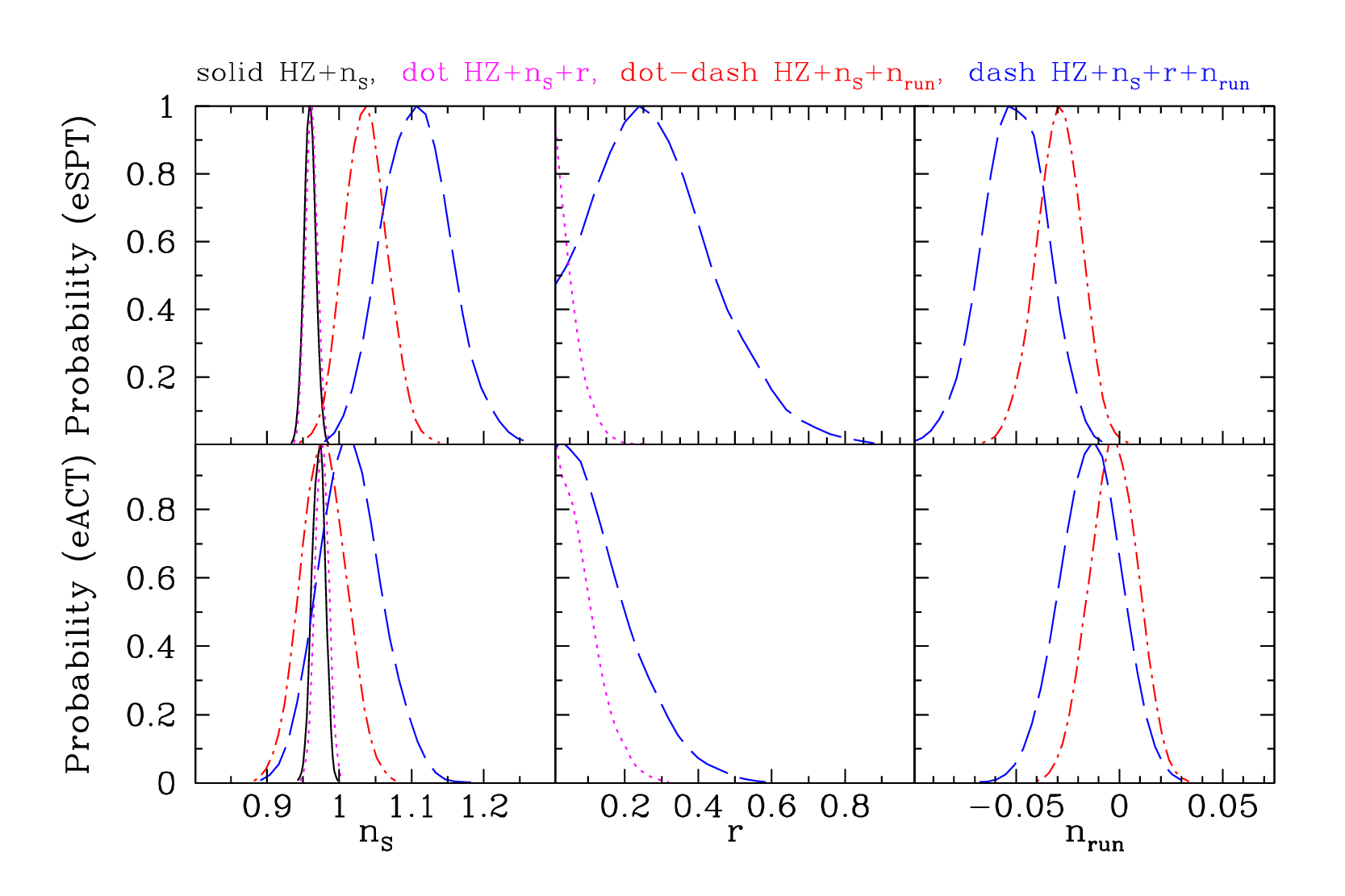} \\
\includegraphics[width=0.32\linewidth]{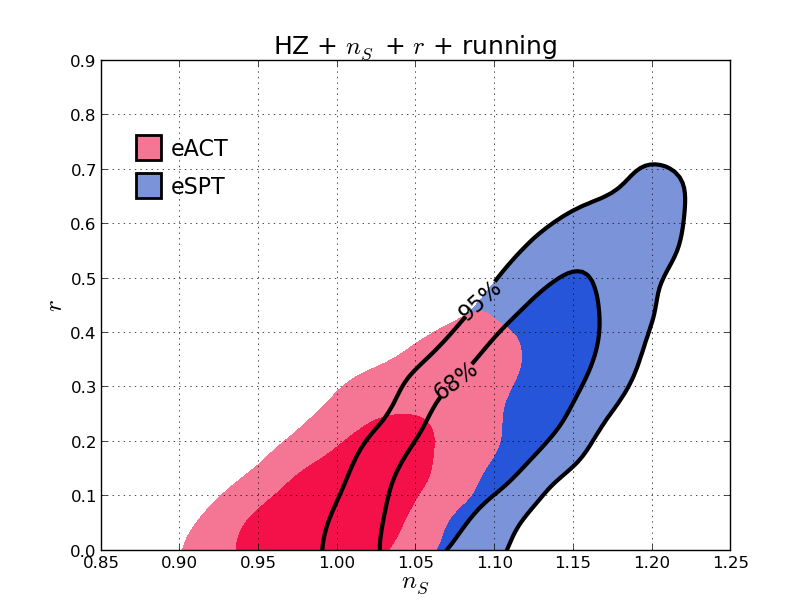}
\includegraphics[width=0.32\linewidth,keepaspectratio]{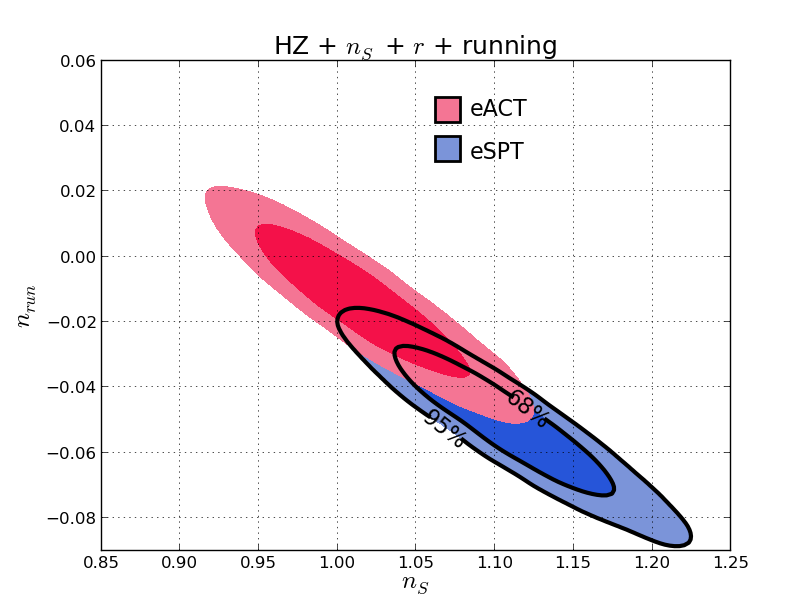}
\includegraphics[width=0.32\linewidth,keepaspectratio]{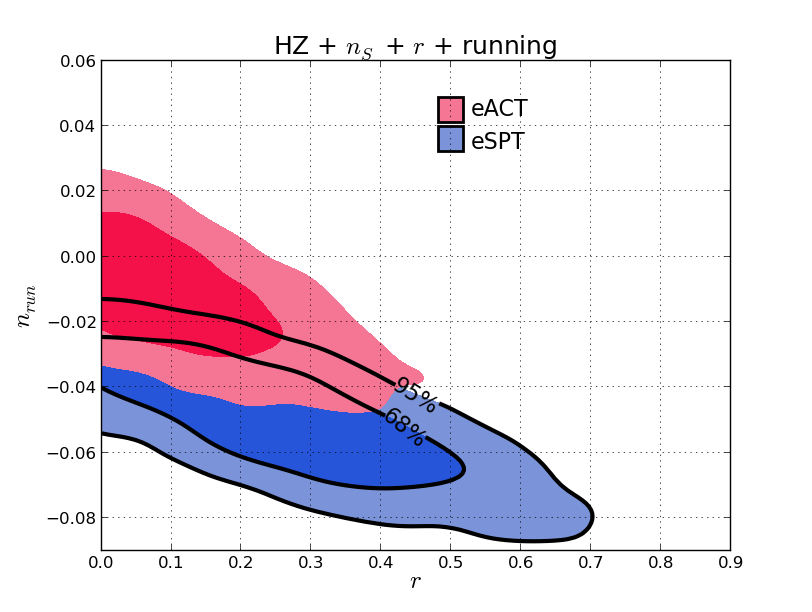}
\caption{One- and two-dimensional posterior probabilities for $n_S$, $r$, and $n_{\mathrm run}$. Upper panel: One-dimensional parameter posteriors
 for the models considered in the text, using the eSPT (top row) and eACT (bottom row) datasest. Lower panel: Two-dimensional posteriors for the HZ+$n_s$+$r$+running case. Dark- and light-shaded regions correspond to 68 and 95\% credible intervals, respectivey. }
\label{fig:inflpost}
\end{center}
\end{figure*}

For the eSPT dataset, allowing one additional parameter, $n_S$, to vary results in change in $\chi^2$ of $\Delta\chi^2\equiv \left(- 2\log\mathcal{L}\right)-\left(- 2\log\mathcal{L}_{HZ}\right)=-28.7$.  The one-dimensional probability distribution for $n_S$ with the eSPT dataset is shown in Fig.\ \ref{fig:inflpost}.  For the eACT dataset, allowing one additional parameter, $n_S$, to vary results in $\Delta\chi^2\equiv \left(- 2\log\mathcal{L}\right)-\left(- 2\log\mathcal{L}_{HZ}\right)=-9.7$.  The one-dimensional probability distribution for $n_S$ with the eACT dataset is also shown in Fig.\ \ref{fig:inflpost}.  

\begin{figure*}
\begin{center}
\includegraphics[width=0.45\linewidth,keepaspectratio]{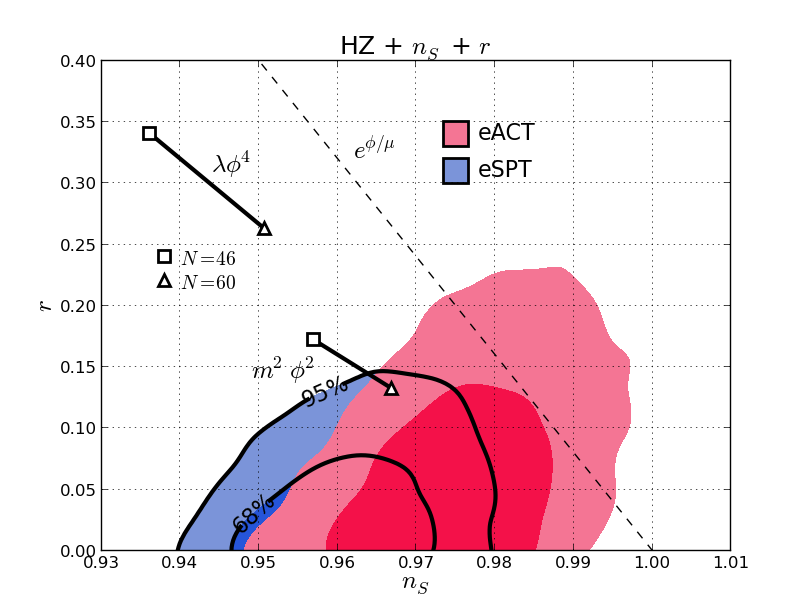}
\includegraphics[width=0.45\linewidth,keepaspectratio]{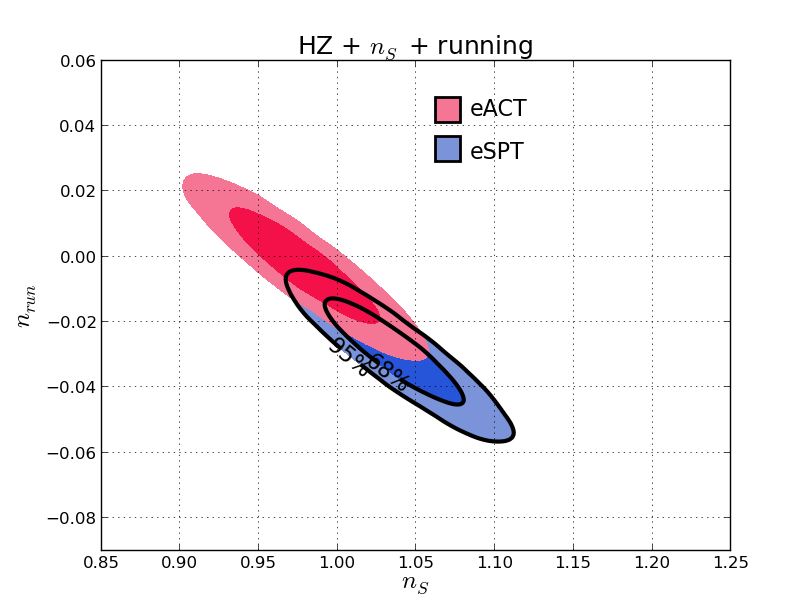}\caption{Two-dimensional probability in the $n_S$ vs.\ $r$ plane for the HZ $+n_S+r$ model in the left panel, and the HZ $+n_S+n_{run}$ model in the right panel.}
\label{fig:inflnr}
\end{center}
\end{figure*}

If we allow other parameters describing the perturbation spectra to vary, such as $\nrun$ and $r$, there are different indications from the different datasets.  Let us first consider the eSPT dataset. 

The natural parameter space for constraining simple slow-roll inflation models is to include the the tensor/scalar ratio $r$ in addition to spectral tilt $n_S$. Two-dimensional contours for $n_S$ vs.\ $r$ are shown in Fig.\ \ref{fig:inflnr}, along with the predictions of three simple slow-roll models. For the eSPT dataset, allowing $r$ to vary in addition to allowing $n_S$ to vary results in a very marginal decrease in $\chi^2$ of $-0.9$ compared to a model just allowing $n_S$ to vary.  Hence, the data do not seem to call for the additional variable $r$.   However, the situation is quite different if we allow a running of the scalar spectral index, $\nrun\neq0$, either keeping $r=0$ or allowing $r$ to vary.  Adding one additional parameter, $\nrun$, results in $\Delta\chi^2=-36.6$ compared to the reference HZ model, which corresponds to a change in $\chi^2$ of $-7.9$ compared to the HZ $+n_S$ model.  If we allow \textit{both} $r$ and $\nrun$ to vary (in addition to allowing $n_S$ to vary) there is a gain of $\Delta\chi^2=-37.5$ compared to the reference HZ model, or a change in $\chi^2$ of $-8.8$ compared to the HZ $+n_S$ model. The eSPT dataset strongly prefers a running of the scalar spectral index.  The one-dimensional probability distributions for $n_S$, $r$ and $\nrun$ with the eSPT dataset are shown in Fig.\ \ref{fig:inflpost}.  Two-dimensional contours of $r$ vs.\ $n_S$, $\nrun$ vs.\ $n_S$ and $\nrun$ vs.\ $r$ are also shown in Fig.\ \ref{fig:inflpost}.

The eACT dataset also prefers a scalar spectral index different from unity.  Recall that adding one additional parameter $n_S$ results in a decrease in $\chi^2$ compared to the reference HZ model of $\Delta\chi^2=-9.7$.  If we then allow one additional parameter, either $r$ or $\nrun$, there is only a very marginal change in $\chi^2$ beyond the HZ + $n_S$ model.  Even allowing both additional parameters $\nrun$ and $r$ again results in a very marginal decrease in $\chi^2$ at the expense of two additional parameters.  The one-dimensional probability distributions for $n_S$, $r$ and $\nrun$ with the eACT dataset are also shown in Fig.\ \ref{fig:inflpost}, and the two-dimensional contours of $r$ vs.\ $n_S$, $\nrun$ vs.\ $n_S$ and $\nrun$ vs.\ $r$ are also shown in Fig.\ \ref{fig:inflpost}.

\begin{figure}
\begin{center}
\includegraphics[width=0.95\linewidth,keepaspectratio]{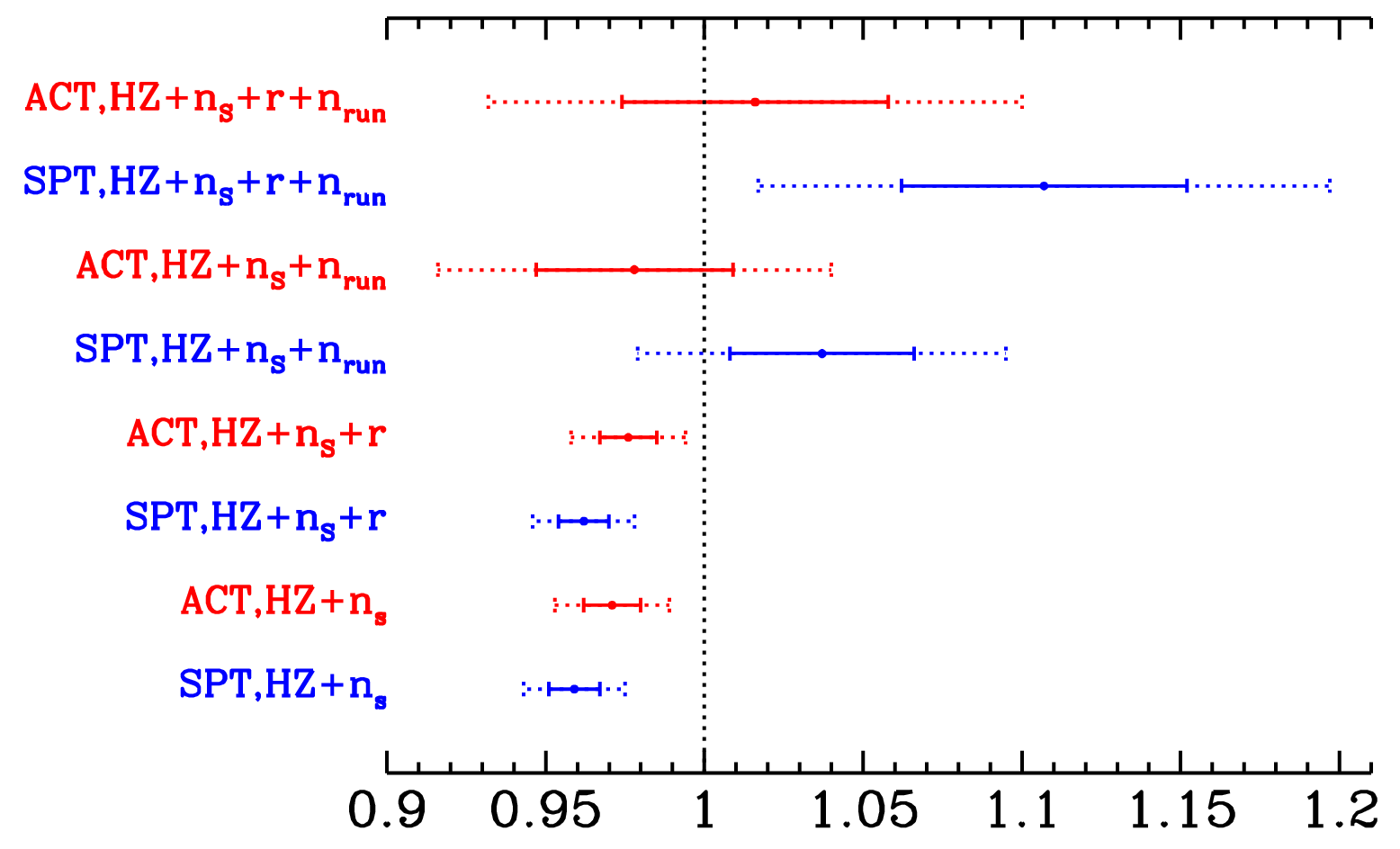} \\[0.4cm]
\includegraphics[width=0.95\linewidth,keepaspectratio]{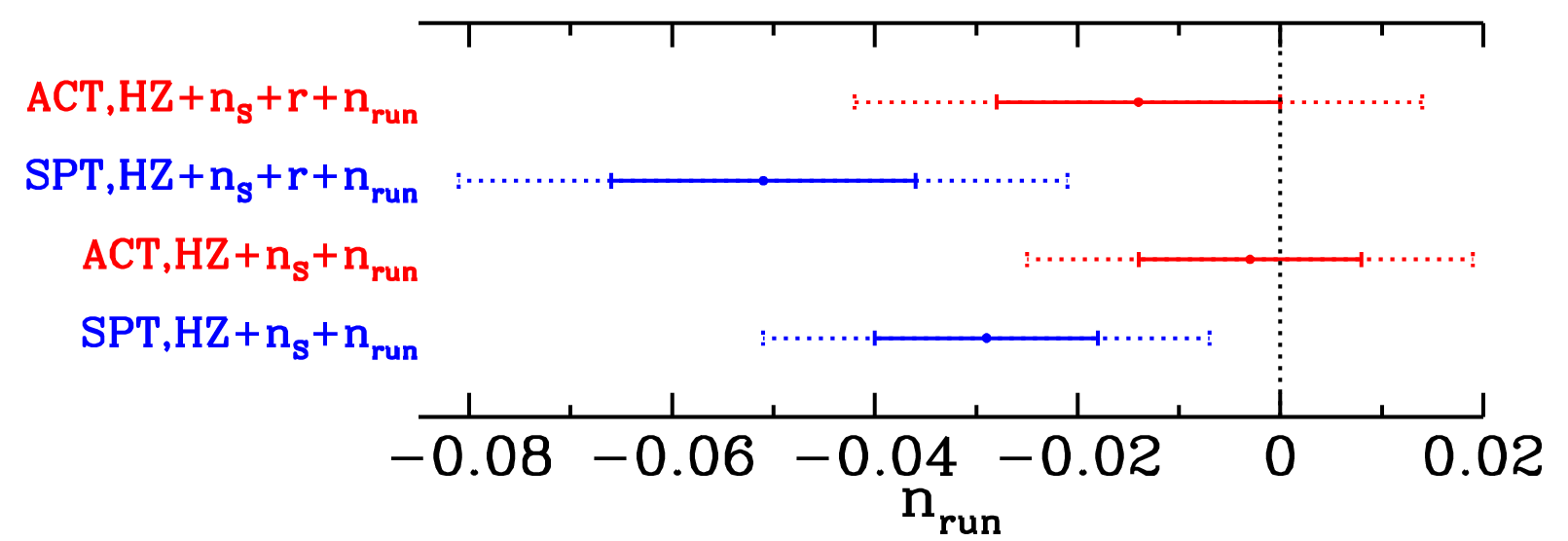}
\caption{Comparing the constraints on $n_s$ (top panel) and $n_\mathrm{run}$ (bottom panel) for different model/datasets combinations. The solid and dashed bars denote 1- and 2-$\sigma$ constraints, respectively .}
\label{fig:ns_nrun_whisk}
\end{center}
\end{figure}

We summarize our findings with respect to $n_s$ and $n_{\mathrm{run}}$ in Fig.\ \ref{fig:ns_nrun_whisk}, where we compare the constraints on these
parameters for the different model/dataset combinations considered in the paper. It is clear from this figure that the tension between the two datasets 
increases when the model complexity is also increased. Moreover, as discussed above in the context of the goodness-of-fit of the various models, we 
also notice that the results of parameter estimation from eACT are more stable, with respect to eSPT, to the increase of the complexity of the model.

Our conclusion is that the eSPT and eACT datasets are not consistent, as long as inflation-motivated extensions to the minimal model are concerned.  While both call for a scalar spectral index different than unity, the eSPT dataset seems to be better described by a more complicated perturbation spectrum than just a scalar spectrum of constant spectral index.  On the other hand, the eACT dataset seems to be well described by a constant scalar spectral index (slightly less than unity), and does not seem to require additional complexity.

\subsection{Extensions of the neutrino sector \label{sec:nuquovadis}}

\begin{table*}[htb!]
\caption{Augmenting the minimal Harrison-Zel'dovich cosmological model through new neutrino physics.  Listed are posterior means for the cosmological parameters from the indicated datasets (errors refer to 68\% credible intervals, unless otherwise stated).}
\begin{ruledtabular}
\label{tab:Stew_nu}
\footnotetext [1] {eV}
\footnotetext [2] {km s$^{-1}$ Mpc$^{-1}$}
\footnotetext [3]{When comparing to the $\chi^2$ values reported e.g. in the WMAP9 paper \cite{wmap9}, it should be taken into account that we use a pixel based likelihood at low $l$s instead than the gibbs-based likelihood.}
\footnotetext [4] {$\Delta\chi^2\equiv \left(- 2\log\mathcal{L}\right)-\left(- 2\log\mathcal{L}_{HZ}\right)$}
\footnotetext [5] {95\% c.l.}
\begin{tabular}{c|c|c|cccc}
\multirow{2}{*}{Dataset} & \multirow{2}{*}{Parameter} & Reference Model \phantom{X}& \multicolumn{3}{c}{Neutrino-Motivated Extensions} \\
& & HZ & HZ $+\neff$ & HZ $+m_\nu$ & HZ $+\neff+m_\nu$ \\ \hline
\multirow{10}{*}{eSPT} 
& $100\,\Omega_b h^2$ 		
& $2.331\pm0.025$&$2.311\pm0.024$&$2.330\pm0.024$& $ 2.332\pm0.037$\\
& $\Omega_c h^2$
& $0.1148\pm0.0017$&$0.1394\pm0.0057$&$0.1100\pm0.0023$&$0.1315\pm0.0057$\\
& $100\, \theta$                		
& $1.0430\pm0.0009$&$1.0404\pm0.0010$&$1.0434\pm0.0009$&$1.0412\pm0.0011$\\
& \phantom{X}$\log[10^{10} A_S]$\phantom{X}			
& $ 3.12\pm0.03$&$3.15\pm0.03$&$3.12\pm0.03$& $3.14\pm0.03$\\
& $\tau$ 		                  		
& $0.096\pm0.013$&$0.085\pm0.012$&$0.103\pm0.014$&$0.095\pm0.014$\\
& $\neff$					
& $\equiv3.046$&$4.26\pm0.26$&$\equiv3.046$&$4.45\pm0.32$\\
& $\sum m_{\nu}$\ \ $^{(a)}$ 	
& $\equiv0$&$\equiv0$&$0.39\pm0.14$&$0.96\pm0.53$\\
& $H_0$ \ \ $^{(b)}$
& $71.33\pm0.65$&$75.5\pm1.1$&$69.82\pm0.76$& $ 74.0\pm1.2$ \\
& $- 2\log\mathcal{L}$ \ \ $^{(c)}$
& $7653.4$ &$7625.9$&$7645.3$&$7617.1$ \\
& $\Delta\chi^2$\ \ $^{(d)}$
& $\equiv0$ &$-27.5$&$-8.1$ & $-36.3$ \\
\hline
\multirow{10}{*}{eACT} 
& $100\,\Omega_b h^2$ 		
& $2.356\pm0.027$& $2.332\pm0.029$&$2.358\pm0.029$&$2.337\pm0.029$\\
& $\Omega_c h^2$
& $0.1163\pm0.0021$& $0.1318\pm0.0057$&$0.1156\pm0.0021$&$0.1296\pm0.0057$\\
& $100\, \theta$                		
& $1.0416\pm0.0016$&$1.0382\pm0.0020$&$1.0421\pm0.0016$&$1.0387\pm0.0020$\\
& \phantom{X}$\log[10^{10} A_S]$\phantom{X}			
& $3.14\pm0.03$&$3.16\pm0.03$&$3.13\pm0.03$& $3.15\pm0.03$\\
& $\tau$ 		                  		
& $0.102\pm0.014$&$0.097\pm0.014$&$0.105\pm0.015$&$0.099\pm0.014$\\
& $\neff$					
& $\equiv3.046$&$3.88\pm0.28$&$\equiv3.046$&$3.80\pm0.28$\\
& $\sum m_{\nu}$\ \ $^{(a)}$ 	
& $\equiv0$&$\equiv0$&$0.24\pm0.15$&$<0.46$\ \ $^{(e)}$\\
& $H_0$ \ \ $^{(b)}$
& $70.50\pm0.71$&$73.2\pm1.1$&$69.82\pm0.79$& $ 72.4\pm1.2$ \\
& $- 2\log\mathcal{L}$ \ \ $^{(c)}$       
& $7617.9$ &$7609.7$&$7616.7$&$7609.2$ \\
& $\Delta\chi^2$\ \ $^{(d)}$
& $\equiv0$ &$-8.2$&$-1.2$ & $-8.7$ \\
\end{tabular}
\end{ruledtabular}
\end{table*}

We now repeat the analysis presented in previous subsection but now considering the possibility of an extra effective neutrino number and including neutrino masses.

The constraints from the eSPT and eACT dataset are in Table \ref{tab:Stew_nu}.  For both datasets, adding the additional parameter $\neff$ greatly improves the fit.  In fact, allowing $\neff$ improves the fits of both eSPT and eACT by about as much as allowing the spectral index to vary from unity.  

However, allowing the neutrino mass to vary, we again again obtain different indications from the two datasets. 
For the SPT dataset, adding a neutrino mass improves the $\chi^2$ by $-8.1$ if $\neff$ is kept fixed and by $-8.8$ if it is allowed to vary.
For the ACT dataset, on the contrary, the goodness of fit improves only marginally 
(at the price of one additional parameter) by allowing a non-zero neutrino mass, independently of whether $\neff$ is fixed or not.

\begin{figure*}
\begin{center}
\includegraphics[width=0.9\linewidth,keepaspectratio]{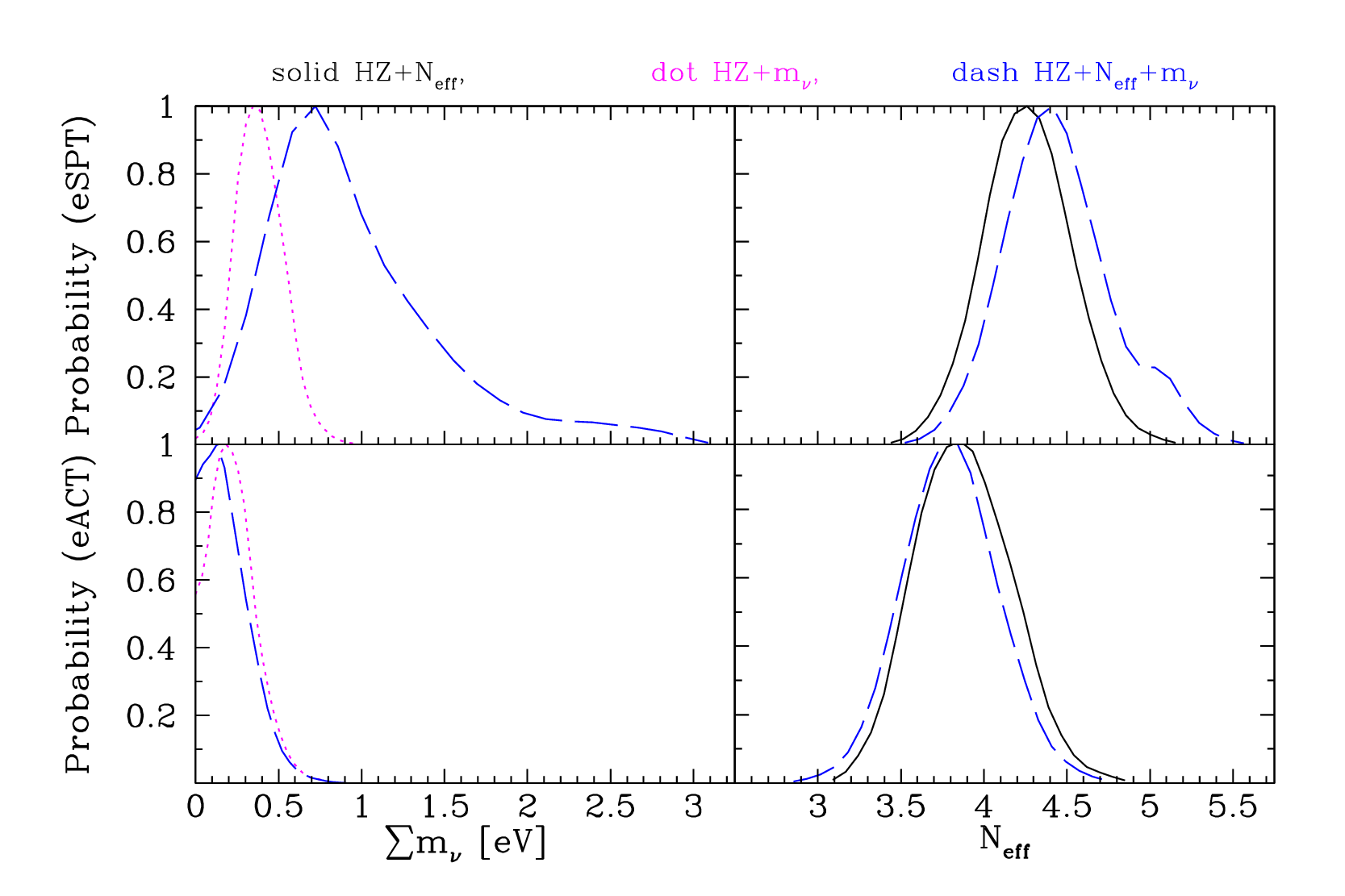} \\
\includegraphics[width=0.5\linewidth,keepaspectratio]{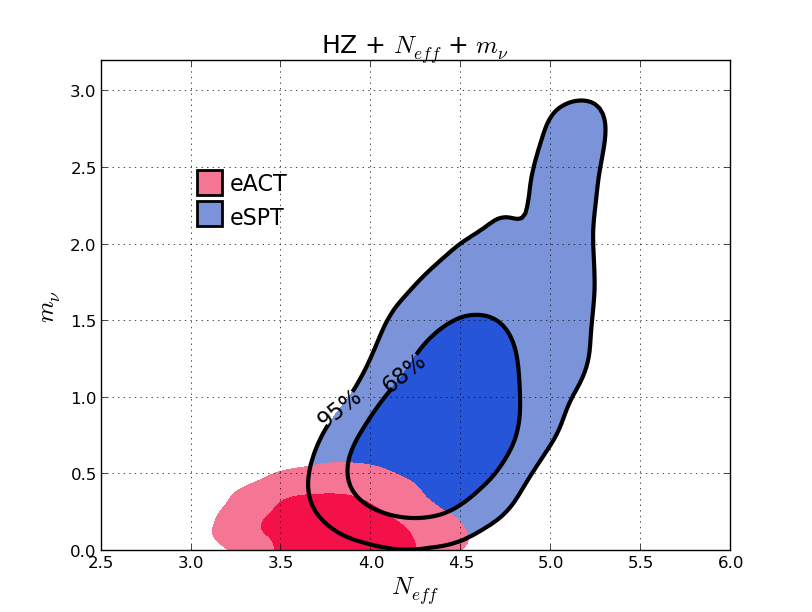}
\caption{One- and two-dimensional posterior probabilities for the eACT and eSPT data for the parameters $\neff$ and $m_\nu$.}
\label{fig:nupost}
\end{center}
\end{figure*}

\begin{figure}
\begin{center}
\includegraphics[width=0.95\linewidth,keepaspectratio]{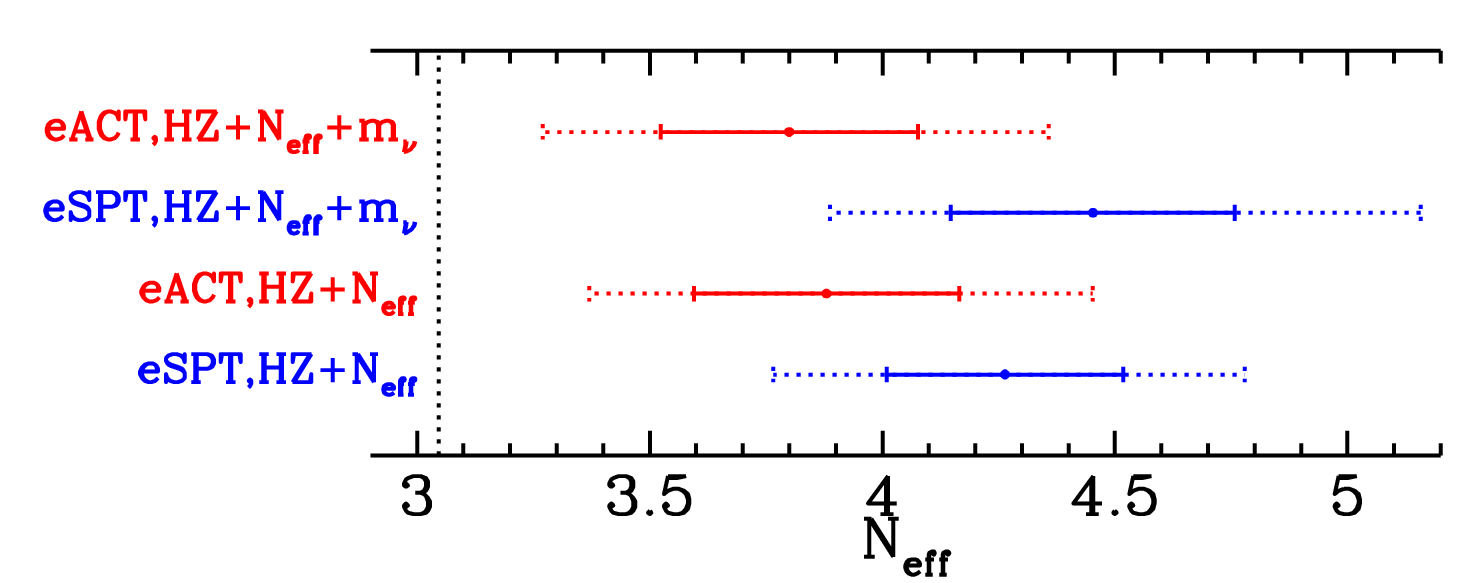} \\[0.4cm]
\includegraphics[width=0.95\linewidth,keepaspectratio]{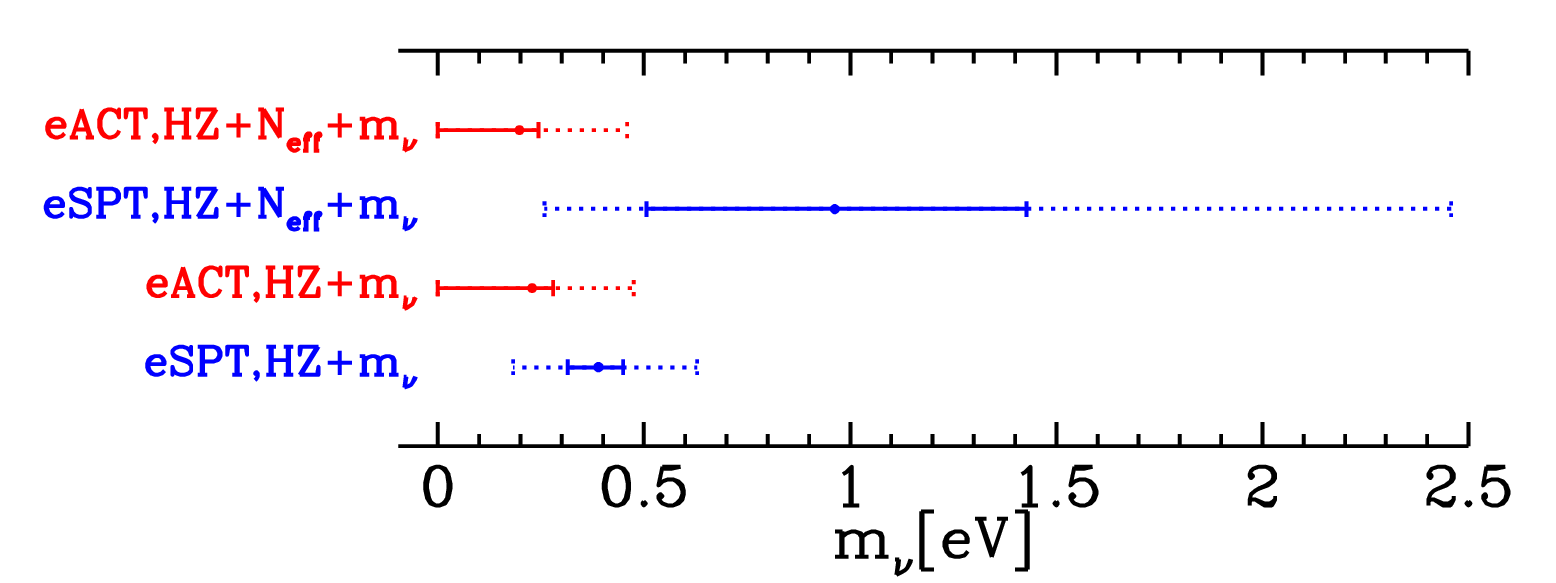}
\caption{Comparing the constraints on $n_s$ (top panel) and $n_\mathrm{run}$ (bottom panel) for different model/datasets combinations. The solid and dashed bars denote 1- and 2-$\sigma$ constraints, respectively .}
\label{fig:nu_whisk}
\end{center}
\end{figure}

In Fig.\ \ref{fig:nu_whisk} we compare the constraints on $\neff$ and $\sum m_\nu$ for the different model/dataset combinations considered in the paper. 
Again we see the same trend observed in the case of the spectrum parameters, namely that the values estimated from the two datasets tend to diverge
as new parameters are added, and that the values estimated from eACT are more stable than those estimated from eSPT when the complexity of the model is increased.

\section{Directions for New Physics \label{sec:quovadis}}

Using the two data combinations described here, models with either primordial perturbations beyond the HZ model or additional light degrees of freedom provide a much better fit than the HZ model.  Thus, cosmological data point to some interesting new physics.  Unfortunately, the direction is unclear.  

Probably the most dramatic explanation would be additional light degrees of freedom:  It would be very surprising if there is a new light species beyond the standard model of particle physics (as we have emphasized, it need not be extra neutrino species, although we parameterize them as such).  

For the eACT dataset, just adding a tilt to the scalar spectrum seems to be all that is demanded of the data.  This would tell us something about inflation, but there are a large number of inflation models that can give a slightly red spectrum.

For the eSPT dataset however, the data seems to demand more than simply a tilt to the scalar spectrum. 
A much improved fit can be obtained by allowing the possibility of a large running of the scalar spectrum.  The running could be so large as to have a large impact in inflation model building and  call in doubt the simple slow-roll approximation.  Alternatively, as data     seem to indicate  a non-power-law scalar spectral index with a very large variation of the spectral index, one might invoke models where the flattening of the
inflaton potential is obtained through the inclusion of large quantum corrections in the mass parameter \cite{sed} which result
in large variation of the spectral index with the scale. Another class of models which allow for a large negative running are models in which inflation occurs near an inflection point of the potential, where the third derivative $V'''$ of the potential is substantial, and the higher-order slow roll parameter ${}^2\lambda$ is comparable to the lower-order parameters $\epsilon$ and $\eta$. Inflection point inflation models have been argued, \textit{e.g.} in Ref. \cite{McAllister:2012am}, to be characteristic of inflation on the string landscape. 

\section{Conclusions \label{sec:conclusions}}

We analyzed the recently released Atacama Cosmology Telescope (ACT) and South Pole Telescope (SPT) data in combination with the Wilkinson Microwave Anisotropy Probe 9-year data (WMAP9), the Sloan Digital Sky Survey Data Release 9, the WiggleZ large-scale structure data, and the Hubble Space Telescope determination of the Hubble parameter (HST). We tested these data against two cosmological scenarios: (1) a scale-invariant, purely scalar ``Harrison-Zel'dovich'' (HZ) power spectrum with the addition of parameters motivated by inflationary cosmology, tilt $n_S$, nonzero tensor/scalar ratio $r$, and running of the spectral index $n_S$, and (2) the HZ power spectrum with a nonstandard effective neutrino number $N_{\mathrm eff}$ and/or neutrino mass $m_\nu$. We find that both the extended ACT data (eACT) and the extended SPT data (eSPT) favor extensions to the simple HZ model to at least 95\% confidence. 

In the case of the inflation-motivated extensions to HZ, both eACT and eSPT favor a deviation from a scale-invariant power spectrum with ``red'' tilt, $n_S < 1$, and neither show any evidence for a nonzero tensor/scalar ratio. The eACT data are consistent with negligible running of the spectral index, as predicted by simple slow-roll inflation models. The eACT data are consistent at the 95\% confidence level with simple chaotic inflation $V\left(\phi\right) = m^2 \phi^2$, and with power-law inflation, $V\left(\phi\right) \propto \exp\left(\phi / \mu\right)$, as well as ``small-field'' models predicting negligible tensors and $n_S < 1$. The eSPT data, however, are inconsistent with a purely power-law power spectrum, favoring negative running of the spectral index $n_{\mathrm run} = -0.029 \pm 0.011$ in the case with a prior of $r = 0$, and  $n_{\mathrm run} = -0.051 \pm 0.015$ in the case where $r \neq 0$ is allowed. While the eSPT data are not in disagreement with the most general possible single-field inflation models, they are in significant conflict with slow-roll models predicting $n_{run} \ll n_{S}$. The eACT data are consistent with such models. 

In the case of extensions to HZ involving additional light degrees of freedom, eACT and eSPT again produce qualitatively different constraints. Both the eACT and eSPT data favor additional light degrees of freedom, with $N_{\mathrm eff} = 3.88 \pm 0.28$ for eACT, and $N_{\mathrm eff} = 4.26 \pm 0.26$ for eSPT (with a prior of $m_\nu = 0$). The eACT and eSPT data differ, however, with respect to nonzero neutrino masses. The eACT data are consistent at 95\% with zero neutrino mass, with $\sum{m_\nu} = 0.24 \pm 0.15$ eV (with a prior of $N_{\mathrm eff} \equiv 3.04$), and  $\sum{m_\nu} < 0.46$ eV (with $N_{\mathrm eff} \neq 3.04$). The eSPT data favor nonzero neutrino mass, with $\sum{m_\nu} = 0.39 \pm 0.14$ eV (with a prior of $N_{\mathrm eff} \equiv 3.04$), and  $\sum{m_\nu} = 0.96 \pm 0.53$ eV (with $N_{\mathrm eff} \neq 3.04$). 

In either scenario, HZ + inflation or HZ + neutrinos, considering the ACT and SPT data separately results in qualitatively different conclusions about extensions to a standard scale-invariant $\Lambda$+Cold Dark Matter concordance cosmology, a tension which is not evident when considering combined constraints from ACT and SPT.

\acknowledgments
A.R.\ is supported by the Swiss National Science Foundation (SNSF), project ``The non-Gaussian Universe'' (project number: 200021140236). WHK is funded by the U.S. National Science Foundation grant NSF-PHY-1066278. The work of ML has been supported by Ministero dell'Istruzione, dell'Universit\`a e della Ricerca (MIUR) through the PRIN grant ``Galactic and extragalactic polarized microwave emission'' (contract number PRIN 2009XZ54H2-002). We would like to thank Luca Amendola for useful discussion.


\end{document}